\documentclass[twocolumn,showpacs,preprintnumbers,amsmath,amssymb]{revtex4}

\begin{document}
\draft

\title{Importance of individual scattering matrix elements at Fano resonances}

\author{P. Singha Deo}
\address{ S. N. Bose National Centre for Basic Sciences, JD Block,
Sector III, Salt Lake City, Kolkata 98, India.}
\date{\today}

\begin{abstract}
Single particle resonances in quantum wires are generally Fano resonances.
In case of Fano resonances, the scattering phase shift in some channels
show sharp phase drops
and that in the other channels do not. 
Phase shift in a particular channel can be measured and can yield
information about the integrated charge localized around the scatterer.
This paper tries to analyze if some channels are more informative
than the others, so that an experimentalist can measure the phase shift
in only those channels.

\end{abstract}


\maketitle

\section {INTRODUCTION}

Miniaturization of devices is an ever pressing demand. It is also
obvious that to sustain miniaturization, we have to use systems
that are so small that
the laws of physics is determined by Quantum Mechanics.
So understanding the properties of so called mesoscopic
systems is also very demanding \cite{dat}. Circuits of future devices
will be made up of quantum wires. Junctions in such circuits
also has to be understood from Quantum mechanics. Such a simple
junction is what is refered to as a three prong potential \cite{gan}
and it is schematically shown in Fig. 1. There are several
papers that explore the importance of such a junction in a 
mesoscopic system \cite{deo96,pas}.

While bound states are important to understand the
thermodynamic properties of a sample \cite{akk}, the scattering states
are important to understand the transport properties \cite{das}. 
However,
it is not always necessary to solve the bound states and the scattering
states separately. Both these states are solutions of the same
Hamiltonian but with different boundary conditions. It is possible
to solve only the scattering states and thereby also understand the 
bound states as an analytical continuation of the scattering states.
At the heart of this principle is the Friedel sum rule \cite{swa,tan}
that states that for non-polarizable leads \cite{swa}
\begin{equation}
{d \theta_f \over dE} 
\approx \pi[\rho(E)-\rho_0(E)]
\end{equation}
The correction terms are small in WKB regime \cite{swa}. Here 
\begin{equation}
\theta_f = {d \over dE} {1 \over 2i} log Det[S], 
\end{equation}
$E$ is
the incident electron energy, $S$ is the scattering matrix, $\rho(E)$
is the density of states (DOS) at energy E in presence of scattering,
$\rho_0(E)$ is DOS in absence of scattering.
$\theta_f$ is refered to as the Friedel phase. The RHS of Eq. 1 is therefore
related to  the integrated charge localized around the scatterer and this can
be determined from the $S$ matrix.

A number of experiments \cite{sch, kob1, kob2} 
show that the phase of a particular scattering
matrix element can be measured. Yeyati and Buttiker \cite{yey1} first
proposed that such phases can be understood from the Friedel sum rule (FSR).
It was later found that transmission zeroes play a special role \cite{deo96,
deo2,yey2} and change our understanding of the FSR.

\section{The problem}

Normally, in mesoscopic systems,
$S$ is a large but finite
matrix, and often all the matrix elements are
not completely independent. So, it may not be necessary to evaluate
all the elements of $S$, in order to calculate $\theta_f$. Moreover,
in mesoscopic systems, it is possible to measure a particular matrix
element of $S$. One can measure the phase as well as the amplitude
of the matrix element \cite{sch}.
For a large dimensional $S$ matrix, it may not be possible to
measure all the elements of $S$.
So, it is important to know what kind of physical informations can
be obtained from the phase of a particular matrix element
and which are the important matrix elements. For example,
in case of a one dimension (1D) system,
the scattering matrix is 2x2, and \cite{har}
\begin{equation}
{d \over dE} arg(T) = {d \theta_f \over dE}
\end{equation}
where $T$ is the transmission amplitude and $arg(T)= Arctan{Im[T] \over
Re[T]}$. Situation is complicated in quasi-1D where a single
transverse mode is populated and the scattering matrix is 2x2.
This is because of the presence of discontinuous changes
in scattering phase shifts at Fano resonances \cite{deo96}.
For such systems it was shown \cite{tan}
\begin{equation}
{d \over dE} arg(T)= {d \theta_f \over dE} \pm \pi \delta(E-E_0)
\end{equation}
Here, $E_0$ is the Fermi energy at which $T=0$. $T=0$ occur because of
Fano resonance and is a general feature of quasi-1D \cite{deo2}. Situation is
further complicated in multichannel quasi-1D where the dimensionality
of $S$ matrix is more than 2x2. 
In case of a multichannel Fano resonance, there are sharp
continuous phase drops at the minima of certain scattering probabilities,
while not so for others \cite{swa}.
For example, in case of the $\delta$ function potential in a
multichannel quantum wire, there are sharp drops in
$arg(T_{mm})$ versus incident energy, when
$|T_{mm}|^2$ is minimum. But, this does not happen
for $arg(T_{mn})$ or $arg(R_{mn})$ or
$arg(R_{mm})$.
There it was shown \cite{swa}
\begin{equation}
{d \over dE} arg(R_{mn}) = 
{d \over dE} arg(R_{mm}) = 
{d \over dE} arg(T_{mn}) = 
{d \theta_f \over dE} 
\end{equation}
where $m \ne n$, $R_{mn}$ is reflection amplitude
from $m$th transverse channel to
$n$th transverse channel,
$R_{mm}$ is reflection amplitude
from $m$th transverse channel to
$m$th transverse channel and
$T_{mn}$ is transmission amplitude
from $m$th transverse channel to
$n$th transverse channel.
Eq. 5 also applies to the potentials studied in Ref. \cite{tan}
but eq. 4 does not apply to the multi channel wire
with a $\delta$ function potential.

Which means charge transfer is determined by the energy derivatives of
those $arg(S_{\alpha \beta})$ that do not show a drop.
Their derivatives are identical to the derivatives of the
Friedel phase $\theta_f$.
However, delta function
potential, as well as the potentials studied in Ref. \cite{tan} are all
point scatterers and it is not known if this feature is general.
This motivates us to study an extended potential that can
exhibit multi channel Fano resonance.

So in this paper we study the system shown in Fig. 1, as it is
an extended potential. We shall show that it can exhibit
multichannel Fano resonance. It is a simple model, that
allows analytical understanding and allows us to check if
the $S$ matrix elements whose phase do not show a negative slope
carry the information about the DOS.
Besides, the potential
considered here has certain subtleties. A different kind of subtlety exists
\cite{swa} for a potential in quasi-1D, 
namely there is an exponentially decaying
local density of states (LDOS) 
in the leads due to the evanescent modes. One has to be careful
about interpreting
what is a non-polarizable lead that normally means that the integrated
LDOS in the leads are not be considered in (1). But in quasi 1D
it is not possible to separate the LDOS in the leads and the LDOS
in the sample. The exponentially decaying LDOS in the leads are to be
included in (1) but the non-decaying LDOS in the leads
should not be included in (1) \cite{swa}.
As a result, the correction terms to Eq. (1)
are negligible in WKB regime is
no longer a correct statement \cite{swa}.
In the present case, the subtlety arises with the interpretation of
$\rho_0$ as will be explained in section IV.

\section{The scattering solution}

Consider the geometry, schematically shown in Fig. 1. 
The thin lines are semi-infinite 1D quantum wires (that successfully models
a quasi 1D quantum wire with a single transverse mode populated)
with quantum mechanical potential $V=0$,
while the thick lines are quantum wires with a potential $V \ne 0$.
An incident electron is shown by the arrow head and the quantum
mechanical wave functions in the different regions I, II, III,
IV, V and VI are written below.
\begin{equation}
\psi_I=e^{ik(x-l_1)} + R_{11} e^{-ik(x-l_1)}
\end{equation}
\begin{equation}
\psi_{II}=A e^{iqx} + B e^{-iqx}
\end{equation}
\begin{equation}
\psi_{III}=C e^{iqy} + D e^{-iqy}
\end{equation}
\begin{equation}
\psi_{IV}=F e^{iqz} + G e^{-iqz}
\end{equation}
\begin{equation}
\psi_V=T_{13} e^{ik(z-l_3)}
\end{equation}
\begin{equation}
\psi_{VI}=T_{12} e^{ik(y-l_2)}
\end{equation}
$l_1$, $l_2$ and $l_3$ are defined in Fig. 1. These wave functions
have to be continuous and current has to be conserved at the 
junctions. From this we get a set 9 linear equations from which
we can solve the 9 unknown coefficients that are $A, B, C, D,
F, G, R_{11}, T_{12}$ and $T_{13}$.
Similarly, we have to solve the scattering problem when the
incident electron is from the top, in order to obtain
$R_{22}, T_{21}$ etc.
Thus we can determine the scattering
matrix.  It is a three channel problem and the
scattering matrix is 3x3 as shown below. 
 \begin{equation}
 S = \left(\begin{array}{ccc}
 \displaystyle R_{11} &
 \displaystyle T_{12} &
 \displaystyle T_{13} \\
 \displaystyle T_{21} &
 \displaystyle R_{22} &
 \displaystyle T_{23} \\
 \displaystyle T_{31} &
 \displaystyle T_{32} &
 \displaystyle R_{33}
 \end{array}\right) \hspace{.2cm}.
 \end{equation}

It is known that \cite{dat}
$$\rho(E)=-{1 \over \pi} Im Tr[G^{ret}(r, r', E)]=$$ 
$${\sum_{k,l}}
\delta(E-E_{k,l}) \int_{-\infty}^\infty dr | \psi_{k,l}(r)|^2
=\rho^{(W)}(E) +$$
$${2 \over hv}\int_{- \infty} ^{l_1} dx + 
{2 \over hv}\int^{ \infty} _{l_2} dy + 
{2 \over hv}\int^{ \infty} _{l_3} dz +$$ 
$${2 |R_{11}| \over h v} \int_{- \infty} ^{ l_1} cos(2k x + \eta_1) dx
+{2 |R_{22}| \over h v} \int_{- \infty} ^{ l_2} cos(2k y + \eta_2) dy$$
\begin{equation}
+{2 |R_{33}| \over h v} \int_{- \infty} ^{ l_3} cos(2k z + \eta_3) dz
\end{equation}
Here 
$G^{ret}(r,r',E)$ is the retarded Green's function and
$\rho^{(W)}(E)$ is the integrated local DOS in the
thick region of Fig. 1. Allowed modes in the system are denoted
by momentum index $k=\sqrt{{2m \over \hbar^2} (E-V)}$ 
and $l$ can take three values (1, 2 and 3),
corresponding to incident electron from three possible
directions. Here $m$ is electronic mass and $v$ is ${\hbar k \over m}$.
$\eta_1= - arg(R_{11})$ and so on.
Here
\begin{equation}
\rho^{(W)}(E)=\rho_1^{(W)}(E)+ \rho_2^{(W)}(E)+ \rho_3^{(W)}(E)
= {\sum_l} \rho_l^{(W)} (E)
\end{equation}
where
$$\rho_1^{(W)}(E)=
{1 \over hv} [\int_{l_1}^0 |A e^{i q x} + B e^{-i q x}|^2 dx +$$
\begin{equation}
\int_0^{l_2} |C e^{i q y} + D e^{-i q y}|^2 dy +
\int_0^{l_3} |F e^{i q z} + G e^{-i q z}|^2 dz ]
\end{equation}
That is $\rho_1^{(W)}(E)$ is calculated with the incident electron
from the left.
$\rho_2^{(W)}(E)$ is calculated with the incident electron from the top.
And $\rho_3^{(W)}(E)$ is calculated with the incident electron from
the right.
Substituting eq. 13 in eq. 1
and considering the mesoscopic leads to be unpolarizable
(Which means the last 3 terms on the RHS of eq. 13 are to be ignored. 
Actually even if one does not want to invoke the non-polarizability
of the leads, one can evaluate those indefinite integrals and see
that these terms are negligibly small in the regime of
interest, which is the WKB regime. This is the regime where
transport occurs) we get
\begin{equation}
{d \theta_f \over dE} \approx \pi[\rho^{(W)}(E)-\rho^{(W)}_0(E)]
\end{equation}

\section {Elimination of $\rho^{(W)}_0(E)$}

In case of the three prong potential, it is not
possible to realize the state of ``absence of scattering". Even
if we make the potential $V$ in Fig. 1, equal to 0, we get $T_{12}
=T_{13}=2/3$
and $R_{11}=-1/3$ for $l_1=l_2=l_3$. Even for unequal $l_1$, $l_2$ and
$l_3$, there is always some reflection and scattering, in absence of
the potential. This is often encountered in case of scattering
by topological defects where there is no continuous way of going
to the defect free regime. So first we have to eliminate $\rho^{(W)}_0(E)$,
i.e., the DOS in absence of scattering. This is dealt with in this
section.

One can say that when $V=0$, then
\begin{equation}
{d \theta'_f \over dE} \approx \pi [\rho^{(W')}(E)-\rho^{(W)}_0(E)]
\end{equation}
Here, $W'$ stands for the thick region in Fig. I with $V$ set to 0.
$\theta '$ is calculated from the $S$ matrix when $V=0$.
$\rho_0^{(W)}(E)$ is the DOS of the unknown system, that is the
defect free version of the scattering system being studied.
$V=0$ situation yields a $S$ matrix given below.
 \begin{equation}
 S' = \left(\begin{array}{ccc}
 \displaystyle -(1/3)e^{2ikl_1} &
 \displaystyle (2/3)e^{2ikl_2} &
 \displaystyle (2/3)e^{2ikl_3} \\
 \displaystyle (2/3)e^{2ikl_1} &
 \displaystyle -(1/3)e^{2ikl_2} &
 \displaystyle (2/3)e^{2ikl_3} \\
 \displaystyle (2/3)e^{2ikl_1} &
 \displaystyle (2/3)e^{2ikl_2} &
 \displaystyle -(1/3)e^{2ikl_3}
 \end{array}\right) \hspace{.2cm}.
 \end{equation}
From this it can be proved that 
\begin{equation}
{d \theta'_f \over dE}= {d \over dE} (-{1 \over 2i} log Det[S']) =
{2l_1 \over hv} + {2l_2 \over hv} + {2l_3 \over hv} \approx \rho^{(W')}(E)
\end{equation}
Similarly, when $V\ne 0$,
\begin{equation}
{d \theta_f \over dE} \approx \pi [\rho^{(W)}(E)-\rho^{(W)}_0(E)]
\end{equation}
From 17 and 20, we find that
\begin{equation}
{d \theta_f \over dE} - {d \theta'_f \over dE}  \approx  \pi 
[\rho^{(W)}(E) - \rho^{(W')}(E)]
\end{equation}
Thus using 19 we get
\begin{equation}
{d \theta_f \over dE} \approx \pi \rho^{(W)(E)}
\end{equation}
Or
\begin{equation}
{d \theta_f \over dk} \approx \hbar v \rho^{(W)}(k)
\end{equation}
With the choice of wavefunctions in (6) - (11), this is also what we get for
potential scattering. This is because the S-matrix elements of the
three prong potential for $V \rightarrow 0$, 
does not have any non-trivial energy
dependence (energy dependence is same as that of a free particle). But this
method of elimination can be applied to complicated junctions where we will
get situations not encountered with potential scattering.  

\section{Results and discussions}

We first show that the three prong potential can exhibit
multichannel Fano resonance.
In Fig. 2, we show that $arg(T_{13})$ drops sharply
when $|T_{13}|^2$ minimizes. In Fig. 3, we show that
$arg(R_{11})$ drops sharply when $|R_{11}|^2$ minimizes.
These are typical signatures of Fano resonances \cite{swa}.
These phase drops can be sharper than any scale present
in the system as shown in Figs. 2 and 3.
These phase shifts can be detected experimentally and it is
important to investigate the physical informations that can
be obtained from such phase shifts.
In the following we will show that $arg(T_{12})$
does not show any drop and also calculate the DOS
explicitly to check if DOS or Friedel
phase is related to $arg(T_{12})$.
Our previous experience, as discussed in section II,
suggests that this could be the case.

In Fig. 4, we plot ${d \over d(kl)} arg(T_{12})$ as a function
of $kl$. It shows sharp peaks and is positive everywhere, implying that
unlike in $arg(T_{13})$ or $arg(R_{11})$, there are no drops or negative slopes
in  $arg(T_{12})$. Also ${d \over d(kl)} arg(T_{12})$ is very close
to ${d \theta_f \over d(kl)}$ and $\hbar v \rho^{(W)}(kl)$. Here
$l$ is taken to be the unit of length. However, unlike the case
of the $\delta$ function potential in Q1D,  ${d \over d(kl)} arg(T_{12})$
is not identical to ${d \theta_f \over d(kl)}$. They are indeed very
close to each other and identical for practical purposes
when the Fano resonances are long lived. 
However, for Fano resonances
with short life times, there are quantitative differences,
although there is qualitative similarity, as shown in Fig. 5.

\section{Conclusions}

Fano resonances occur very frequently in mesoscopic systems.
Not much is known about
scattering phase shifts for Fano resonances. 
We show that the three prong potential is a simple model
that gives multichannel Fano resonance.
Unlike, previously studied potentials that show Fano resonances,
this is an extended potential. In case of Fano resonances, 
scattering phase shifts in only
some particular channels show sharp phase drops while others do not.
The channels that do not show the phase drops seem to be the
more informative channels and hence very special channels. It seems that 
DOS is related to the
scattering phase shifts of these special channels. As, this is exact
for point scatterers, we were tempted to check it for an
extended scatterer that exhibit Fano resonance.
The three prong potential studied shows that this is true
when the Fano resonances are long lived. 
In all the experiments so far \cite{sch, kob1, kob2} one has encountered
long lived Fano resonances where the phase drops are very sharp.
As the life time
of the resonances decrease, the energy derivative of
the scattering phase shifts in these special channels and 
integrated charge inside the scatterer start deviating
from each other. However, they are qualitatively
similar. The phase shift in the channels that exhibit
the phase drops do not give any information about the
DOS, qualitatively or quantitatively, for long
lived resonances or short lived resonances. So far, experiments \cite{sch,
kob1, kob2}
and theories \cite{tan,yey2} mostly focused on the phase shifts
of these non-informative channels. We hope that this will
give some clues to future works to find a mathematical proof
of this fact.
Such a proof should be consistent with the features observed
in the three prong potential and in the $\delta$ function
potential in a quantum wire.

\centerline{\bf Figure Captions}

\noindent Fig. 1. A schematic diagram of the system studied. Three
semi-infinite
quantum wires meet at a point B. In the thin regions the quantum
mechanical potential $V$ is zero but in the thick regions it is not zero.
The arrow shows the direction of propagation of an incident electron.

\noindent Fig. 2. $arg(T_{13})$ [solid curve] 
decreases sharply but continuously, when $|T_{13}|^2$ [dotted 
curve] minimizes. This is a signature of multichannel Fano resonance.
$l$ is the unit of length.

\noindent Fig. 3. $arg(R_{11})$ [dotted curve] 
decreases sharply but continuously, when $|R_{11}|^2$ [solid 
curve] minimizes. This is a signature of multichannel Fano resonance.
$l$ is the unit of length.

\noindent Fig. 4. ${d \over d(kl)}arg(T_{12})$ [solid curve]
is positive everywhere,
implying that there are no negative slopes in $arg(T_{12})$.
${d \over d(kl)}arg(T_{12})$ [solid curve] and 
${d \theta_f \over d(kl)}$ [dotted curve]
are very close to each other, implying that  $arg(T_{12})$
should carry all the information about DOS.
Explicit calculations of $\hbar v \rho^{(W)}(kl)$ [dashed]
curve confirms this. $l$ is the unit of length.

\noindent Fig. 5. Here we plot the same things as in Fig. 4.
Difference is that the resonances are short lived as compared
to that in Fig. 4. The peaks in $\hbar v \rho^{(W)}(kl)$
are broader sallow as compared to that in Fig. 4. ${d \over d(kl)}arg(T_{12})$
is qualitatively similar to ${d \theta_f \over d(kl)}$ or 
$\hbar v \rho^{(W)}(kl)$. There are quantitative differences.

\end{document}